\documentclass{jetpl}
\usepackage{epsfig,bm,amsfonts,amssymb}
\twocolumn

\def\be{\begin{equation}}
\def\ee{\end{equation}}
\def\x{{\bf x}}
\def\y{{\bf y}}
\def\k{{\bf k}}
\def\p{{\bf p}}
\def\ds{\displaystyle}
\def\vp{\varphi}

\title{Chiral symmetry and excited baryons}
\rtitle{Chiral symmetry and excited baryons}

\author{A.\,V.\,Nefediev}
\address{Institute of Theoretical and Experimental Physics,
117218, B.Cheremushkinskaya 25, Moscow, Russia}
\author{J.\,E.\,F.\,T.\,Ribeiro}
\address{Centro de F\'\i sica das Interac\c c\~oes
Fundamentais (CFIF),Departamento de F\'\i sica, Instituto Superior
T\'ecnico, Av. Rovisco Pais, 049-001,
Universidade T\'ecnica de Lisboa, Lisbon, Portugal}
\author{Adam\,P.\,Szczepaniak}
\address{Department of Physics and Nuclear Theory Center
Indiana University, Bloomington, IN 47405, USA }
\rauthor{A.\,V.\,Nefediev, J.\,E.\,F.\,T.\,Ribeiro, and Adam\,P.\,Szczepaniak}

\abstract{An approach to baryons in the framework of the microscopic Generalized Nambu--Jona-Lasinio chiral potential quark model is considered and
quite general arguments are given in favor of effective restoration of chiral symmetry in excited baryons.}


\begin{document}
\maketitle

Chiral symmetry breaking and confinement of color charges are the two most prominent phenomena taking place in QCD. In spite of the rather
long time that has elapsed since the appearance of QCD as the theory of strong interactions, and the many theoretical efforts thereof, the
fundamental underlying mechanisms of confinement and its interplay with chiral symmetry breaking cannot be derived directly from
first principles. In such circumstances, quark models can be used as a source of information on various properties
of hadrons. In particular, the problems related to chiral symmetry are well suited to be treated with the help of the Generalized
Nambu--Jona-Lasinio (GNJL) model \cite{Orsay,Lisbon}. In this model, the spontaneous breaking of chiral symmetry  happens as a result of
the vacuum condensation of ${}^3P_0$ quark--antiquark pairs, an energetically favorable process as it leads to the decrease of the vacuum
energy. The new BCS vacuum (as opposed to the trivial empty vacuum) is not chirally invariant and thence all hadronic states built on top
of this vacuum should display the same lack of chiral invariance. With the exception of the notorious case of the pion (as its mass would
be higher than its actual physical mass, by $300\div 400$ MeV, were it not for the mechanism of chiral symmetry spontaneous breaking) this
lack of chiral invariance is quite visible with pairs of low--lying hadronic partner states, with opposite parities, being well split in
mass. However, this mass split should become weaker as the hadronic excitation number increases. This phenomenon, known as the asymptotic
effective restoration of chiral symmetry \cite{G1} (see also the recent review \cite{Grev} and references therein) receives a natural explanation
based on the fact that quantum fluctuations should progressively disappear as we climb the excitation number staircase of hadronic states
\cite{G4,GNR}. In other words, in the limit of very large excitation numbers, we expect these very excited hadrons to become semiclassical
objects. A pattern of chiral symmetry restoration in excited heavy--light mesons in the framework of GNJL model was studied in detail in
Refs.~\cite{KaNeRi,GW,ANERAS} and its connection to the classical limit of the model was discussed in Ref.~\cite{GNR}.

In this paper, we give general
arguments concerning the aforementioned effective restoration of chiral symmetry in the spectrum of excited baryons. As in the case of mesons,
chiral restoration in the baryonic sector is an inevitable consequence of the large--momentum behavior of the chiral angle --- the
function which naturally quantifies chiral symmetry breaking.

\begin{figure}[t]
\begin{center}
\epsfig{file=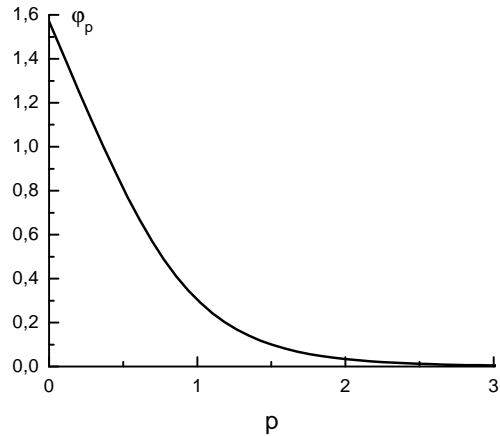,width=6.5cm} \caption{A typical profile of the chiral angle --- solution to the mass-gap equation
(\ref{mg}) which defines the broken BCS vacuum. The momentum $p$ is measured
in the units of strength of the potential $V(|\x-\y|)$.}\label{vpfig}
\end{center}
\end{figure}

For this work, let us consider the simplest Hamiltonian containing
the ladder Dyson--Schwinger machinery for chiral symmetry \cite{Orsay,Lisbon} --- in any case, the results
presented here do not depend on the kernel choice (the superscripts $i,j$ stands for the quark flavor),
\begin{eqnarray}\label{Hamiltonian}
H&=&\sum_{i=u,d }\int d^3x\psi^{\dag}_i({\x})\left(-i{\bm \alpha} {\bm \bigtriangledown}+\beta m_i\right)\psi_i({\x})\nonumber\\[-3mm]
\\[-3mm]
&+&\frac12\sum_{i,j=u,d}\int d^3 xd^3 y\;J^a_{i\mu}({\x})K^{ab}_{\mu\nu}({\x}-{\y})J^b_{j\nu}({\y}),\nonumber
\end{eqnarray}
with
$J_{i\mu }^{a}(x)=\bar{\psi}_{i\alpha}(x)\gamma_{\mu }\left(\frac{\lambda^{a}}{2}\right)^\alpha_\beta\psi_i^\beta(x)$,
$K_{\mu\nu}^{ab}(x-y)=\delta^{ab}K_{\mu\nu}(|\x-\y|)$. The quark field $\psi_i^\alpha(\x)$ is given by
\be
\psi^\alpha_i(\x)=\sum_{s=\uparrow,\downarrow}\int\frac{d^3p}{(2\pi)^3}e^{i\p\x}[b^\alpha_{ips}u_s(\p)+d^{\alpha\dagger}_{ips} v_s(-\p)],
\label{bandd}
\ee
which is an abstraction, independent of any particular choice of the representation for the quark creation/annihilation operators.
If we choose the trivial, empty, chirally symmetric vacuum $|0\rangle_0$ to be the ground state, then the operators $b^\dagger$ and $d^\dagger$
create bare quarks, and the spinors
can be conveniently written as,
\begin{eqnarray}
u_{\rm bare}(\p)&=&\ds\frac{1}{\sqrt{2}}\left[1+({\bm\alpha}\hat{\p})\right]u_0(\p),\nonumber\\[-3mm]
&&\\[-3mm]
v_{\rm bare}(-\p)&=&\ds\frac{1}{\sqrt{2}}\left[1-(\bm{\alpha}\hat{\p})\right]v_0(-\p)\nonumber,
\end{eqnarray}
where $u_0(\p)$ and $v_0(-\p)$ are the rest-frame spinors. However, it was found long ago \cite{Orsay,Lisbon} that the true vacuum
state of the theory (\ref{Hamiltonian}) is given by chirally nonsymmetric state $|0\rangle=e^{Q_0^\dagger-Q_0}|0\rangle_0$, with
\be
Q_0^\dagger=\frac12\sum_{i=u,d}\int\frac{d^3p}{(2\pi)^3}\vp_p^{(i)} b^\dagger_{i\alpha ps}[\mathfrak{M}_{^3P_0}]_{ss'}d^{\alpha\dagger}_{ips'},
\label{S0}
\ee
and the $^3P_0$ matrix being $\mathfrak{M}_{^3P_0}=({\bm\sigma}\hat{\bf p})i\sigma_2$ \cite{Lisbon}. In Eq.~(\ref{S0}) and in what follows
summation over repeated color and spin indices is understood. The quantity $\vp_p^{(i)}$ is known as the chiral angle
and it defines the distribution of the ${}^3P_0$ quark--antiquark pairs of the given flavor in the vacuum according to their relative
momentum. Dependence of the chiral angle on flavor appears entirely due to different masses $m_i$. In what follows we assume an exact
$SU(2)_f$ symmetry, so that $m_u=m_d=m$ and, as a result, $\vp_p^{(u)}=\vp_p^{(d)}\equiv \vp_p$.
The choice of the profile of the function $\vp_p$ ensures that the vacuum energy is minimal,
\be
\frac{\delta}{\delta\vp_p}\langle 0|H|0\rangle=0,
\label{var}
\ee
which leads to the equation for the chiral angle,
known as the mass-gap equation \cite{Orsay,Lisbon}\footnote{Notice that a more correct definition of the
chiral angle and the mass-gap equation follows from the requirement that anomalous
Bogoliubov terms are missing in the normally ordered Hamiltonian (\ref{Hamiltonian}) \cite{Lisbon}. Indeed, the theory is actually well-defined in terms of
dressed quarks only for a discrete set of chiral angles --- solutions of the mass-gap equation. Any variation of the chiral angle leads not
only to a change in the vacuum energy but also in an unbalanced creation of quark--antiquark pairs (described by anomalous Bogoliubov terms
in the Hamiltonian). The latter effect cannot be taken into account with the simple variational procedure (\ref{var}). Meanwhile, formally, one and the
same mass-gap equation arises from both aforementioned procedures.}. For illustrative purposes we quote here the form the mass-gap
equation for the simplest Lorentz structure of the inter-quark interaction in the Hamiltonian
(\ref{Hamiltonian}) compatible with the requirements of chiral symmetry and confinement,
$K_{\mu\nu}(|\x-\y|)=g_{\mu0}g_{\nu0}V_0(|\x-\y|)$ ($V(|\x-\y|)=C_FV_0(|\x-\y|)$, $C_F$ being the fundamental Casimir
operator):
\be
\begin{array}{lcr}
\ds m\cos\vp_p-p\sin\vp_p&&\\[1mm]
\multicolumn{3}{c}{\ds=\frac{1}{2}\int\frac{d^3k}{(2\pi)^3}V(\p-\k)[\sin\vp_k\cos\vp_p}\\[1mm]
&&\ds-(\hat{\p}\hat{\k})\cos\vp_k\sin\vp_p].
\end{array}
\label{mg}
\ee
The properties of this equation and its solutions for various
forms of the quark kernel, in two and four dimensions, can be found in Ref.~\cite{Orsay,Lisbon,repls,NR0}.
It is convenient to define the chiral angle such that $-\pi/2<\vp_p\leqslant\pi/2$ and $\vp(0)=\pi/2$, $\vp(p\to\infty)\to 0$.
The trivial solution of the
mass-gap Eq.~(\ref{mg}) defines the trivial vacuum $|0\rangle_0$, whereas the nontrivial solution (see Fig.~\ref{vpfig} for the profile of
this solution) defines the BCS chirally nonsymmetric vacuum $|0\rangle$, which is the physical vacuum of the theory. From now on, we
choose to deal with the physical vacuum only, so that the operators $b^\dagger$ and $d^\dagger$ in the definition (\ref{bandd}) create
dressed quarks, that is, bare quarks enshrouded by an entire quark--antiquark cloud appearing as a consequence of quark
selfinteractions. The spinors $u(\p)$ and $v(-\p)$ depend on the chiral angle now and can be written as
\begin{eqnarray}
u(\p)&=&\ds\left[\sqrt{\frac{1+\sin\vp_p}{2}}+
({\bm\alpha}\hat{\p})\sqrt{\frac{1-\sin\vp_p}{2}}\right]u_0(\p),\nonumber\\[-1mm]
&&\label{uandv}\\[-1mm]
v(-\p)&=&\ds\left[\sqrt{\frac{1+\sin\vp_p}{2}}-
(\bm{\alpha}\hat{\p})\sqrt{\frac{1-\sin\vp_p}{2}}\right]v_0(-\p)\nonumber.
\end{eqnarray}

\begin{figure}[t]
\epsfig{file=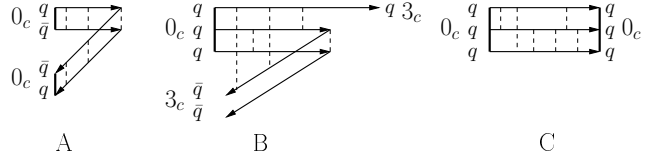,width=8.4cm}
\caption{Diagram A: a typical allowed (singlet $0_c\rightarrow 0_c$)
quark-antiquark pair annihilation transition from mesonic positive-energy
to negative-energy Salpeter amplitudes; Diagram B: a similar pair
annihilation cannot proceed as it will involve non-singlets $3_c$ in color;
Diagram C: a typical diagram pertaining to the Dyson ladder for baryons.}
\label{baryonversusmeson}
\end{figure}

One can see therefore, that the GNJL model gives an explicit microscopic description of the effect of spontaneous breaking of chiral
symmetry. Indeed, in the chiral limit $m=0$ the Hamiltonian (\ref{Hamiltonian}) is invariant under the transformation
$\psi_i\to[\exp\left(i\alpha\gamma _{5}\tau^a/2\right)]^{ij}\psi_j$ (from now onwards $a$ stands for the index of the flavor $\tau$ matrices),
whereas the BCS vacuum $|0\rangle$, contains the chiral condensate
\be
\langle\bar{\psi}\psi\rangle_u=\langle\bar{\psi}\psi\rangle_d=-\frac{N_C}{\pi^2}\int^{\infty}_0 dp\;p^2\sin\vp_p\neq 0,
\ee
and therefore, is not. In this vacuum, the Hamiltonian (\ref{Hamiltonian}), takes a diagonal form
\be
H=E_{\rm vac}+\sum_{i=u,d}\int\frac{d^3 p}{(2\pi)^3}E_p [b_{i\alpha ps}^\dagger b^\alpha_{ips}+d_{ps}^{i\alpha\dagger} d_{i\alpha ps}]+\ldots,
\label{H1}
\ee
where $E_p$ stands for the dressed-quark dispersive law and the ellipsis denotes the terms which are responsible for the
formation of bound states of dressed quarks --- hadrons. In Ref.~\cite{NR0} these terms were considered in detail and a second,
nonlocal Bogoliubov-like transformation was applied in order to diagonalize the Hamiltonian in the mesonic sector of the theory.
Alternatively, one can employ an effective diagrammatic technique with dressed quarks, having quark--antiquark Salpeter amplitudes as
building blocks, in order to derive the Bethe--Salpeter equation --- the bound--state equation for quark--antiquark mesons. This equation was
derived and studied in a vast number of papers --- see, for example, Ref.~\cite{Orsay,Lisbon,NR0,ANERAS}. A key ingredient of this
equation is a two-component w.f. of the meson which describes simultaneously time-forward and time-backward motion of a single, dressed,
quark--antiquark
pair. In other words, quark pair creation--annihilation involves couplings of positive-energy to negative-energy Salpeter
amplitudes. The baryon case is simpler. Due to confinement and to the fact (contrary to the mesonic case) that one
cannot simultaneously annihilate three quarks with a  two body interaction, we do not have such transitions for baryons and,
therefore, to zero order in $N_C$, we are only left with a dressed, positive-energy, three-quark system --- see Fig.~\ref{baryonversusmeson}.

Production of dressed
quark-antiquark pairs is still possible but this is an $N_C$ suppressed effect, responsible, for instance, for the baryon hadronic
decays and virtual hadronic loops. Consideration of such effects goes beyond the scope of the present paper.
Therefore, because of color confinement the w.f. of any baryon can be written as
\be
\Psi_B=\Psi_{\rm color}\otimes\Psi_{\rm flavor}\otimes\Psi_{\rm spin}\otimes\Psi_{\rm space},
\ee
with the color w.f. being antisymmetric, $\Psi_{\rm color}=\frac{1}{3!}\varepsilon_{\alpha\beta\gamma}q^\alpha q^\beta q^\gamma$.
Thus we need to be only concerned with the symmetric $\Psi_{\rm flavor}\otimes\Psi_{\rm spin}\otimes\Psi_{\rm space}$ component of the
baryon w.f. This implies that, in general, the spatial component of the baryon w.f. $\Psi_{\rm space}^{{\cal Y}} ({\bm\rho},
{\bm\lambda)}$
($\bm\rho$ and $\bm\lambda$ are the standard Jacobi coordinates)
must contain all the three-body permutation symmetry components ${\cal Y}$: antisymmetric (${\cal A}$), symmetric (${\cal S}$), mixed type
$F$ (${\cal D}_F$) and, finally, mixed type $D$ (${\cal D}_D$). Recoupling of these spatial w.f.'s with their ${\cal Y}$ flavor--spin counter
parts will produce the completely symmetric w.f. $\Psi_{\rm flavor}\otimes\Psi_{\rm spin}\otimes\Psi_{\rm space}$.

The Noether charge for the global chiral symmetry written in terms of the quark field is:
\be
Q_5^a=\int d^{3}x\;\bar{\psi}_i\gamma_{0}\gamma_{5}\left(\frac{\tau^a}{2}\right)^{ij}\psi_j.
\label{Q5}
\ee

The charge (\ref{Q5}) is to be calculated for the dressed quarks with the help of Eqs.~(\ref{bandd}) and (\ref{uandv}). The result reads:
\be
\begin{array}{lcr}
\multicolumn{2}{l}{\ds Q_5^a=\left(\frac{\tau^a}{2}\right)^{ij}\int\frac{d^3p}{(2\pi)^3}}&\\[5mm]
\multicolumn{3}{c}{\ds\times[\cos\vp_p({\bm\sigma} \hat{{\bf p}})_{ss'}(b_{i\alpha ps}^\dagger b_{jps'}^\alpha
+d_{jps}^{\alpha\dagger} d_{i\alpha ps'})}\\[3mm]
&\multicolumn{2}{r}{\hspace*{1.5cm}\ds+\sin\vp_p(i\sigma_2)_{ss'}(b_{i\alpha ps}^\dagger d^{\alpha\dagger}_{jps'}+d_{i\alpha ps}
b_{jps'}^\alpha)].}
\end{array}
\label{ChiralChar}
\ee

The two terms in square brackets in Eq.~(\ref{ChiralChar}) have
different physical meanings. The second term is known as
the anomalous Bogoliubov term responsible for pion creation. Indeed, in the chiral limit, the axial charge (\ref{ChiralChar}) creates a nontrivial
state, $Q_5^a|0\rangle=|\pi^a\rangle$. Being a Noether charge, the axial charge commutes with the Hamiltonian, $[Q_5^a,H]=0$, which
ensures that the state $|\pi^a\rangle$ is degenerate in energy with the vacuum. This is the Goldstone boson --- the chiral pion \cite{pion}. Its w.f. can be extracted
either from the form of the mass--gap equation or from the Bethe--Salpeter equation for the lowest ${}^1S_0$ bound quark--antiquark state
(see, for example, Ref.~\cite{NR0} for a detailed discussion of the issue). This w.f. is given by $\sin\vp_p$. Finally, the
matrix $i\sigma_2$ in Eq.~(\ref{ChiralChar}) provides the ${}^1S_0$ coupling of the quark with the antiquark.
For quark masses different from zero, the pion acquires a small finite mass. Furthermore,
a simple examination of the form of the first term in square brackets in Eq.~(\ref{ChiralChar}) shows that this operator,
when applied to a given hadronic state, maps it into another hadronic state, with opposite parity, which is ensured by the
matrix $({\bm\sigma} \hat{{\bf p}})$.

Consider the diagonal baryon axial charge operator defined as,
\be
{\cal Q}_5\equiv {\cal Q}_5^3=\sum_{n=1}^3Q_{5n}^3,
\label{Q5bar}
\ee
where index $n$ numerates quarks in the baryons, so that the baryon
total axial charge is given by the sum of three individual charges, one for each quark.

It is obvious from Eq.~(\ref{ChiralChar}) that this
diagonal axial charge acts on baryonic states in a two--fold way, which can be schematically written in the form
($|\pi\rangle$ denotes the neutral pion):
\be
{\cal Q}_5|B\rangle=|B'\rangle+|B\pi\rangle,
\label{hh}
\ee
where, for the sake of simplicity, we suppressed all unnecessary indices.
The states $|B\rangle$ and $|B'\rangle$ have opposite parity, and the relative weight of the two terms on the r.h.s. of Eq.~(\ref{hh})
is defined by the value of the chiral angle $\vp_p$, namely by the value of the $\sin\vp_p$ versus the value of the $\cos\vp_p$.
In case of the maximal symmetry breaking ($\vp_p=\pi/2$),
only the second term on the r.h.s. of Eq.~(\ref{hh}) survives. On the contrary, if chiral symmetry is not broken in the vacuum
($\vp_p\equiv 0$) then the Goldstone does not exist and only the first term survives.
In the latter limit, the following properties of ${\cal Q}_5$ are obvious:
\be\label{paritypairing}
{\cal Q}_5^\dagger={\cal Q}_5,\quad \langle B_2|{\cal Q}_5^2|B_1\rangle\propto\langle B_2|B_1\rangle=\delta_{B_1B_2}.
\ee
Therefore we must have
\be
{\cal Q}_5|B^{\pm}\rangle=G^A_{\pm\raisebox{-0.5mm}{\scriptsize $\mp$}}|B^{\mp}\rangle,
\ee
with $B^{\pm}$ representing baryons with the parity $\pm$ and $G^A_{\pm\raisebox{-0.5mm}{\scriptsize $\mp$}}$ being a $c$-number axial charge.
The latter relation, together with the fact that $[{\cal Q}_5,H]=0$, ensures that, in the chiral limit, the two states
$|B^+\rangle$ and $|B^-\rangle$ must be degenerate in mass. They form a chiral doublet.
This is exactly the regime expected to be approximately realized in highly excited hadrons. Indeed,
in Ref.~\cite{KaNeRi}, the microscopic mechanism for chiral symmetry
restoration in excited hadrons was investigated in detail for the case of heavy--light mesons. The analysis performed in Ref.~\cite{KaNeRi}
applies directly to a heavy--heavy--light baryon with two heavy quarks sitting on top of each other.
In this case, the static particles degrees of
freedom decouple from the system, and the baryon can be described with the light--quark w.f. $\psi$. In agreement with the general
consideration given above, the parity partner of this state has the w.f. $\psi'=({\bm\sigma}\hat{{\bf p}})\psi$ and the two states come out
approximately degenerate in mass in the limit of high orbital or radial excitations, when the momentum of the quark is large and the small
value of the $\sin\vp_p$ (see Fig.~\ref{vpfig}) suppresses the splitting between these two states (see Ref.~\cite{KaNeRi} for the details).
Notice, however, that in the physical world we do not have massless quarks and therefore we expect corrections to this degeneracy arising
from pion loops (this corresponds to proceeding beyond the BCS approximation and taking into account the interaction between dressed quarks).
These corrections are known to be, for the lowest lying baryons, of the order of a few hundred MeV.
Thus the question of chiral restoration lies precisely in whether or not these corrections go to zero for highly excited baryonic states, that is,
whether we can approximate $\langle B' |{\cal Q}_5|B\rangle$ by
\be
\begin{array}{c}
\ds \langle B' |\tilde{\cal Q}_5|B \rangle=\sum_{n=1}^3\left(\frac{\tau_n^3}{2}\right)^{ij}\int\frac{d^3p_n}{(2\pi)^3}
({\bm\sigma}_n \hat{{\bf p}}_n)_{ss'}\\[5mm]
\ds\times\langle B'|b_{i\alpha p_ns}^{(n)\dagger} b_{jp_ns'}^{(n)\alpha}+d_{jp_ns}^{(n)\alpha\dagger} d_{i\alpha p_ns'}^{(n)}|B\rangle.
\end{array}
\label{QA5Chiral}
\ee
where $B$ and $B'$ denote baryons and the index $n$ numerates quarks.

In other words, the issue of chiral restoration in baryons is reduced to the question whether the average quark momentum inside a
sufficiently excited baryon is high enough to have $\sin\vp_p\simeq 0$. In the baryon case and for a generic spatial w.f. with a given
permutation symmetry ${\cal Y}$, we have the following expansion:
\be\label{EquaMaster}
\Psi_B^{{\cal Y}N}=\sum_{C} C^{{\cal Y}N}\sum_{\nu_1\nu_2}D^{{\cal Y}N}_{\nu_1\nu_2}\Phi_{\nu_1}({\bf p}_\lambda)\Phi_{\nu_2}({\bf p}_\rho),
\ee
where ${\bf p}_\lambda$ and ${\bf p}_\rho$ are the momenta conjugate to the Jacobi coordinates ${\bm\lambda}$ and
${\bm\rho}$; $\{\Phi_\nu\}$ is a convenient basis (for example, the
harmonic oscillator one), $N$ stands for the set of quantum numbers, and $D^{{\cal Y}N}$ are
the appropriate coefficients for the particular permutation symmetry ${\cal Y}$ and for the given set of quantum numbers $N$.
The $D^{{\cal Y}N}_{\nu_1\nu_2}$'s obey the normalization condition,
\be\label{DS}
\sum_{\nu_1\nu_2}\left(D^{{\cal Y}N}_{\nu_1\nu_2}\right)^2=1.
\ee
The sum $\sum_{C}$ just reflects the fact that the quarks microscopic interaction, despite being confining and henceforth
infrared divergent, differs from the force with the eigenstates $\{\Phi_\nu\}$. Therefore, all such states should contribute to the baryon
spatial w.f. with weights given by the coefficients $C^{{\cal Y}N}$. Then, Eq.~(\ref{DS}) and orthonormality of baryon states would have
that
\be
\sum_{C}\left(C^{{\cal Y}N}\right)^2=1.
\ee

Then it is clear that for a sufficiently excited baryon, the set $\{C^{{\cal Y}N}\}$ must peak around a given $C^{{\cal Y}N_0}$ ---
with $N_0$ large --- and, as a result, Eq.~(\ref{EquaMaster}) will govern the average value of the quark momenta, which are also large.
Sooner or later this is bound to happen, which results in the chiral angle vanishing. Then two consequences take place:
(i) $\sin\vp_p\to 0$ and pions decouple from this excited baryon (see also Ref.~\cite{GN0} for a detailed discussion of the pion
decoupling from excited hadrons in the framework of GNJL) and (ii) $\cos\vp_p\to 1$, so that the baryon axial charge can be well
approximated by the form given in Eq.~(\ref{QA5Chiral}). As was discussed before, this warrants chiral restoration. Furthermore,
when $\cos\vp_p\neq 1$, the state ${\cal Q}_5|B\rangle$ overlaps with every state of the same parity, therefore entangling
it with states belonging to other chiral multiplets. As soon as chiral symmetry gets restored and $\cos\vp_p=1$, we have zeros
both for the diagonal matrix element, $G^A_{\pm\pm}=\langle B^\pm|{\cal Q}_5|B^\pm\rangle=0$, (because of parity),
and for the matrix elements between states
belonging to different multiplets (because of the w.f. orthogonality). In the meantime, for transitions inside of the same multiplet,
$G^A_{\pm\raisebox{-0.5mm}{\scriptsize $\mp$}}=\langle B^{\mp} |{\cal Q}_5|B^{\pm}\rangle\neq 0$. As seen from Eq.~(\ref{QA5Chiral}),
for highly excited baryons, the off-diagonal axial
charge $G^A_{\pm\raisebox{-0.5mm}{\scriptsize $\mp$}}$ tends to a universal constant independent of the particular baryon multiplet it was calculated
for. By an appropriate rescaling of the baryon axial charge operator, Eq.~(\ref{Q5bar}), this asymptotic value of
$G^A_{\pm\raisebox{-0.5mm}{\scriptsize $\mp$}}$ can be set to unity.
In actuality, since the chiral angle never vanishes identically
and therefore chiral symmetry is never fully restored in the spectrum, we end up with the approximate relations,
\be
G^A_{+-}=G^A_{-+}\simeq 1,\quad G^A_{++}=G^A_{--}\simeq 0,
\ee
which have been obtained microscopically.

We conclude therefore that the GNJL model gives a clear and selfconsistent
pattern of effective chiral symmetry restoration in excited baryons (see the discussion in Ref.~\cite{Grev})
and, what is more, it provides a full {\em microscopic} picture of this phenomenon, which, as a matter of principle, cannot be reproduced
by any naive quark model or approach \cite{GN1}.

As our final remark, we would like to stress that the actual form of the
Hamiltonian (\ref{Hamiltonian}) is of no importance for the above conclusions, provided that ${\cal Q}_5$ is a Noether charge of the
dynamics and we have a nontrivial $\vp_p$.
\bigskip

Useful discussions with L. Ya. Glozman are acknowledged.
Work of A. N. was supported by the Federal Agency for Atomic Energy of Russian Fe\-de\-ration, by the grants
RFFI-05-02-04012-NNIOa, DFG-436 RUS 113/820/0-1(R), by the Federal Programme of
the Russian Ministry of Industry, Science, and Technology No.
40.052.1.1.1112, and by the non-profit ``Dynasty"
foundation and ICFPM. E. R. and A. N. are also supported through the project PTDC/FIS/70843/2006-Fisica.

\end{document}